\documentclass[twoside]{article}
\usepackage{epsfig.sty}
\newcommand{\beq}{\begin{eqnarray}}
\newcommand{\eeq}{\end{eqnarray}}
\begin{document}
\title{Heavy Quark State Production via p-p and O-O Collisions}
\author{Leonard S. Kisslinger\\
Department of Physics, Carnegie Mellon University, Pittsburgh PA 15213 USA.\\
Debasish Das\\ 
Saha Institute of Nuclear Physics, A CI of Homi Bhabha National Institute,\\
1/AF, Bidhan Nagar, Kolkata 700064, INDIA.}
\date{}
\maketitle

\Large

\begin{abstract}

Here we have considered $J/\Psi$ is a normal charmonium meson, while  $\Psi(2S)$ is
a mixed hybrid charmonium meson. Similarly  $\Upsilon(1S)$ and$\Upsilon(2S)$ are
normal upsilon mesons, while $\Upsilon(3S)$ is a mixed hybrid upsilon meson.
We discuss the differential rapidity cross sections for $J/\Psi$, $\Psi(2S)$,
$\Upsilon(1S)$, $\Upsilon(2S)$, $\Upsilon(3S)$ production via p-p, and O-O
collisions at proton-proton energy $\equiv \sqrt{s_{pp}}$= 5.44 TeV.
The rapidity taken for the present study goes from y=-1 to 1.

\end{abstract}
\noindent
\vspace{2mm}

PACS Indices:12.38.Aw,13.60.Le,14.40.Lb,14.40Nd
\vspace{2mm}

\noindent
Keywords: Heavy quark state production; Relativistic heavy ion collisions;
Heavy quark state suppression; Small systems
\vspace{5mm}

\large
\section{Introduction}
\vspace{5mm}

Heavy quark states are among the possible probes of Quark-Gluon Plasma,
QGP\cite{ALICE:2022wpn,brahms05} formed in relativistic heavy-ion collisions.
This article on p-p and O-O collisions is based on the proton-proton
formalism\cite{kd17,lsk09,kd16,klm14} and the color octet
model\cite{cl96,bc96,fl96}. Studies of heavy quark state production in
proton-proton (pp) collisions as a preliminary to relativistic heavy-ion
collisions have been carried out~\cite{bc96}. The ratio of the measured number
of heavy-flavors in heavy ion (A+A) collisions to the expected number in the
absence of nuclear or partonic matter i.e the p+p collisions, is the definition
of nuclear modification factor ($R_{AA}$),  which is suppressed at high
transverse momentum\cite{ALICE:2022wpn}. The elementary degrees of freedom and basic forces at the
shortest distances are explored via small systems~\cite{Das:2023mye}. So a clear picture of the
small systems emerge as a necessity. Also the proposed O+O collisions and data
taken in Xe+Xe collisions assume importance where the system sizes are smaller
than Pb-Pb collisions which are generally taken since the starting of LHC to
study the Quark-Gluon Plasma QGP~\cite{ALICE:2022wpn}. The study of heavy quarks in
small systems like p-p and p-A collisions, can contribute towards, an important
baseline for the A-A collisions as discussed in recent paper~\cite{Das:2023mye}
and references therein.

Prior to the article~\cite{kd16} $\Psi(2S)$ and $\Upsilon(3S)$ suppression in
p-Pb collisions was estimated\cite{lsk16} and reviewed~\cite{lskdd16}.
The ALICE collaboration has studied $J/\Psi$ production~\cite{ALICE:2022wpn,alice18} in Xe-Xe
collisions and recently the possibilities of O-O collisions are being explored
at LHC. Here we have studied them at  $\sqrt{s_{pp}}$ = 5.44 TeV. 

In section 2 Mixed heavy quark hybrid and p-p collisions are discussed.
We have not introduced  Quark-Gluon Plasma (QGP) effects and do not use QGP
in our calculations. In subsection 2.1 $\Psi$ and $\Upsilon$ production via p-p collisions
with $\sqrt{s_{pp}}$ = 5.44 TeV and differential rapidity cross sections for
the production $\Psi(1S)$ and $\Psi(2S)$ via p-p collisions are discussed.
In subsection 2.2, $\Psi$ and $\Upsilon$ production via O-O collisions
with $\sqrt{s_{pp}}$ = 5.44 TeV and the differential rapidity cross sections
are discussed. 

Section 3 gives results and conclusions ending with the References.

\section{Mixed heavy quark hybrid and p-p collisions}

\vspace{4mm}

One possible signal of QGP is the production of heavy quark states in 
relativistic heavy-ion collisions. One indication of the QGP in hybrid
mesons is the energy loss of the quarks, such as a charm quark $c$ and
an anti-charm quark $\bar{c}$ in the hybrid charm meson $\Psi(2S)$.

Studies of heavy quark state production in proton-proton (p-p) collisions
as a preliminary to relativistic heavy-ion collisions have been carried 
out\cite{bc96,fl96}. In \cite{bc96} E. Braaten and Y-Q. Chun derive helicity
decomposition during $J/Psi$ production and the hybrid charm meson $\Psi(2S)$.
\vspace{4mm}

The starting point of the method of QCD sum rules is the correlator
\beq
\label{cor}
       \Pi^A(x) &=&  \langle | T[J_A(x) J_A(0)]|\rangle \; ,
\eeq
with $| \rangle$ the vacuum state and the current $J_A(x)$ creates the 
states with quantum numbers A. For the charmonium states, $J_{c}$ is
\beq
\label{11}
        J_{c} &=& f J_{c \bar{c}} + \sqrt{f^2-1} J_{c \bar{c} g} \; , 
\eeq
where $J_{c \bar{c}}$ creates a normal charmonium state and $J_{c \bar{c} g}$ 
creates a hybrid state with an active gluon.

Using QCD sum rules it was shown that $f \simeq \sqrt{2}$ for the 
$\Psi(2S)$ and $\Upsilon(3S)$ heavy quark meson states and $f \simeq 1.0$ for 
the other charmonium and bottomonium states\cite{lsk09}. Therefore,
\beq
\label{psi}
     |J/\Psi(1S)> &\simeq& |c\bar{c}(1S)>  \\
    |\Psi(2S)> &\simeq& -\sqrt{2} |c\bar{c}(1S)>+\sqrt{2}|c\bar{c}g(2S)> 
\nonumber \; ,
\eeq
with the $\Psi(2S)$ being a mixed hybrid meson.
\vspace{4mm}

Similarly, it was shown that
\beq
\label{up}
  |\Upsilon(3S)> &\simeq& -\sqrt{2} |b\bar{b}(3S)>+\sqrt{2}|b\bar{b}g(3S)> \; ,
\eeq
while
\beq
\label{up1}
  |\Upsilon(1S)> &\simeq&  |b\bar{b}(1S)> \; ,
\eeq
  and
\beq
\label{up2}
  |\Upsilon(2S)> &\simeq&  |b\bar{b}(2S)> \; ,
\eeq
  
An important observation concerning the nature of heavy quark charmonium and
bottomonium states are anomalies: a much larger production of $\Psi(2S)$ in 
high energy collisions than standard model predictions~\cite{cdf}, and the
anomalous production of sigmas in the decay of $\Upsilon(3S)$ to  
$\Upsilon(1S)$~\cite{vogel}. A solution of these anomalies was found in
the mixed hybrid theory~\cite{lsk09,lsk16}. The $\Psi(2S)$ state was found to
be
\vspace{4mm}

\beq
\label{1}
        |\Psi(2S)>&=& \alpha |c\bar{c}(2S)>+\sqrt{1-\alpha^2}|c\bar{c}g(2S)> 
\; ,
\eeq
\vspace{4mm}

where $c$ is a charm quark, and the $\Upsilon(3S)$ state was found to have 
the form
\vspace{4mm}

\beq
\label{2}
    |\Upsilon(3S)>&=& \alpha |b\bar{b}(3S)>+\sqrt{1-\alpha^2}|b\bar{b}g(3S)> 
\; ,
\eeq
\vspace{4mm}

where $b$ is a bottom quark and $\alpha$ = $0.7 \pm 0.1$. This means that 
these states have approximately a 50\% probability of being a standard 
quark-antiquark, $|q\bar{q}>$, meson, and a 50\% probability of a hybrid, 
$|q\bar{q}g>$ with the $|q\bar{q}>$ a color octet and $g$
an active gluon. With a valence gluon it would be natural for these hybrid
states to be produced during the creation of a dense QGP.
\vspace{4mm}

It was shown that using this mixed hybrid theory~\cite{lsk09,lsk16},
upon which the present work is based, that the ratios of $\Psi(2S)/(J/\Psi)$
and $\Upsilon(3S)/\Upsilon(1S)$ agreed with experimental results, while the
standard model for the $\Psi(2S)$ and $\Upsilon(3S)$ did not.


\subsection{$\Psi$ and $\Upsilon$ production via p-p
  collisions with $\sqrt{s_{pp}}$ = 5.44 TeV}
\vspace{4mm}

The differential rapidity cross section for the production of a heavy quark
state $\Phi$ with helicity $\lambda=0$ (for unpolarized  
collisions\cite{klm11}) in the color octet model via Xe-Xe collisions is given 
by\cite{klm14}

\beq
\label{3}
   \frac{d \sigma_{AA\rightarrow \Phi(\lambda=0)}}{dy} &=& 
   R^E_{AA} N^{AA}_{bin}< \frac{d \sigma_{pp\rightarrow \Phi(\lambda=0)}}{dy}>
\; ,
\eeq
\vspace{4mm}

where $R^E_{AA}$ is the product of the nuclear modification factor $R_{AA}$
and $S_{\Phi}$, the dissociation factor after the state $\Phi$ is formed 
(see FIG.3 in Ref\cite{star09}). $R_{AA}$ is defined in Ref\cite{star02} as

\beq
\label{RAA}
  R_{AA}(p)&=& \frac{d^2  N^{AA}_{bin}/dpd\eta}{T_{AA} d^2 \sigma^{NN}/dpd\eta}
\; .
\eeq

For p-p collisions we use  $R^E_{pp}$ =0.005 For O-O collisions we use $R^E_{OO}$ =0.25.

The differential rapidity cross section for p-p collisions 
in terms of $f_g$\cite{klm11}, the gluon distribution function is
\beq
\label{4}
     \frac{d \sigma_{pp\rightarrow \Phi(\lambda=0)}}{dy} &=& 
     A_\Phi \frac{1}{x(y)} f_g(\bar{x}(y),2m)f_g(a/\bar{x}(y),2m) 
\frac{dx}{dy} \; ,
\eeq  
where from Ref\cite{klm11}  $A_\Phi=1.26 \times 10^{-6}$ nb for 
$\Phi$=$J/\Psi$ and $3.4 \times 10^{-8}$ nb for $\Phi$=$\Upsilon(1S)$; and
$a=4 m^2/s =3.6 \times 10^{-7}$ for charmonium and $4.0 \times 10^{-6}$ for 
bottomonium.
\vspace{4mm}

The function $\bar{x}$, the effective parton x in a nucleus (A), is given in
Refs\cite{vitov06},\cite{sharma09}:
\beq
\label{bar}
x(y) &=& 0.5 [\frac{m}{\sqrt{s_{pp}}}(\exp{y}-\exp{(-y)})+
\sqrt{(\frac{m}{\sqrt{s_{pp}}}(\exp{y}-\exp{(-y)}))^2 +4a}] \nonumber \\
\bar{x}(y)&=& (1+\frac{\xi_g^2(A^{1/3}-1)}{Q^2})x(y).
\eeq

\vspace{2mm}

With $\Psi(2S),\Upsilon(3S)$ enhanced by $\pi^2/4$~\cite{klm11} the 
differential rapidity cross sections are shown in the following figures.
The absolute magnitudes are uncertain, and the shapes and relative magnitudes 
are our main predictions. In Eq(\ref{bar}) $m=m_c=$ 1.5 GeV for charmonium
and $m=m_b=$ 5 GeV for bottomonium quarks.
\vspace{2mm}

The differential rapidity cross sections for the
production $\Psi(1S)$ and $\Psi(2S)$ via p-p collisions
are shown in the figures below.
\vspace{2mm}

\begin{figure}[ht]
\begin{center}
\epsfig{file=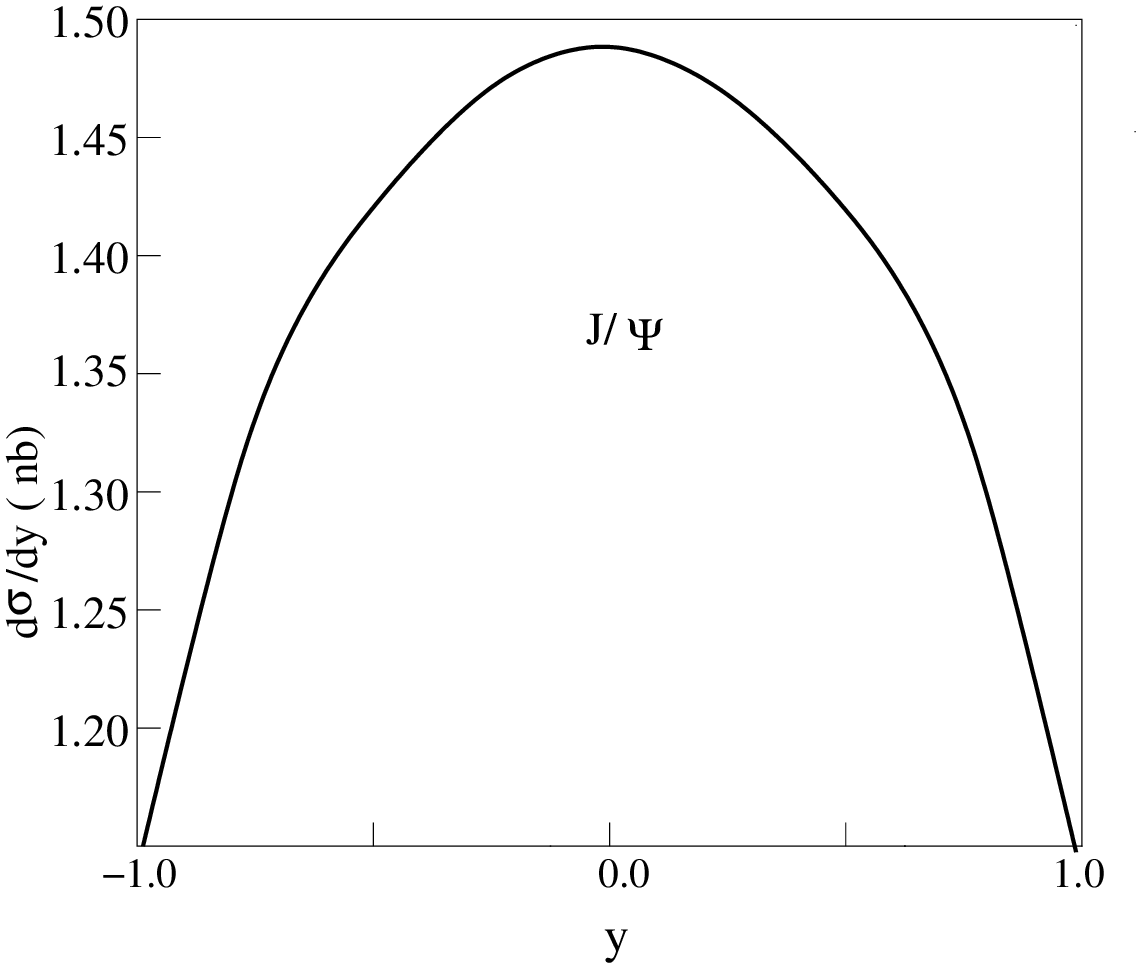,height=10cm,width=15cm}
\caption{d$\sigma$/dy for 2$m_c$=3 GeV, $\sqrt{s_{pp}}$=5.44 TeV via p-p
 collisions producing $J/\Psi$ with $\lambda=0$}  
\end{center}
\end{figure}

\newpage

\begin{figure}[ht]
\begin{center}
\epsfig{file=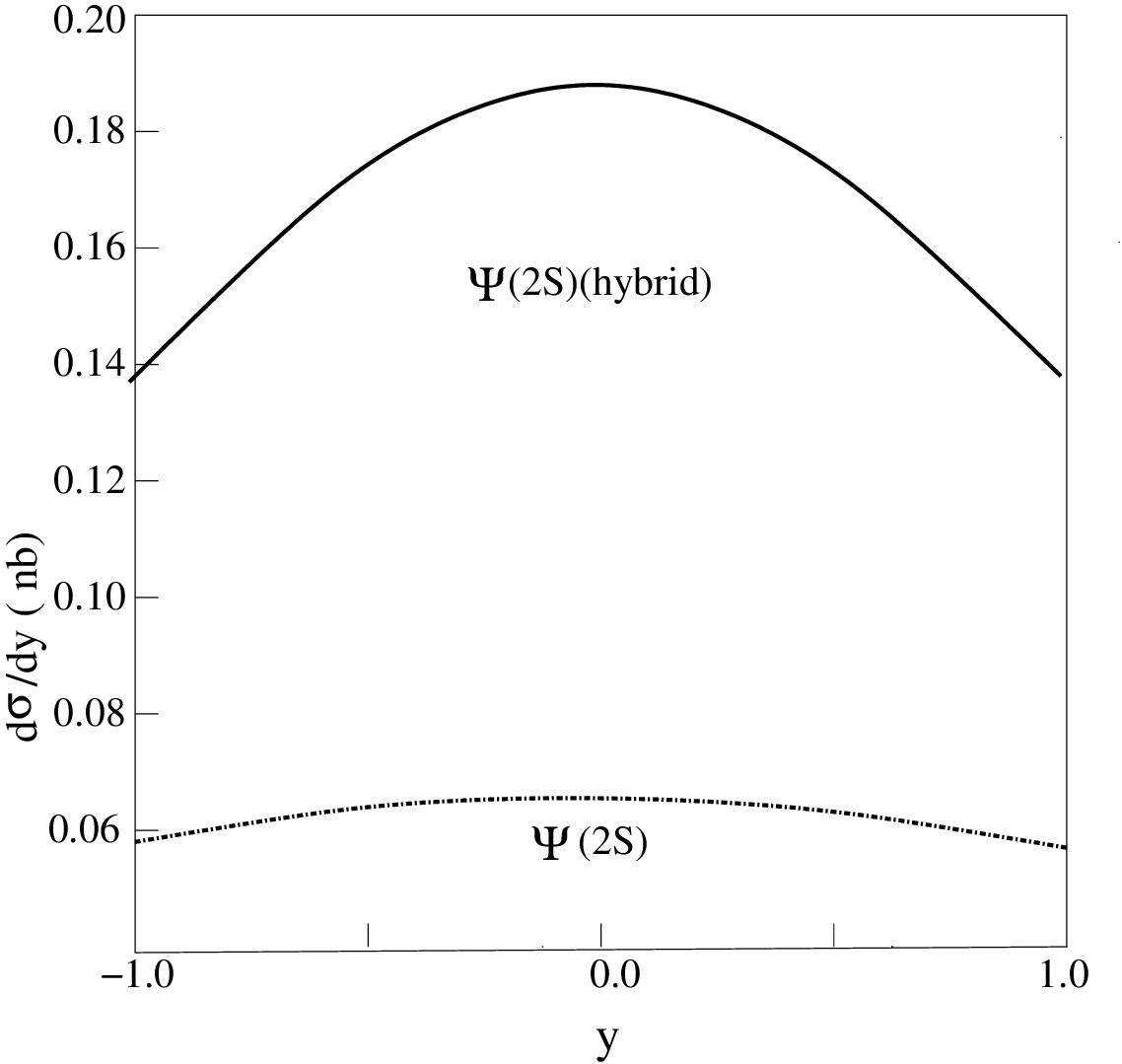,height=7cm,width=8cm}
\caption{d$\sigma$/dy for 2$m_c$=3 GeV, $\sqrt{s_{pp}}$=5.44 TeV via p-p 
collisions producing $\Psi(2S)$, hybrid theory, with $\lambda=0$. The dashed 
curve is for the standard $c\bar{c}$ model.}
\end{center}
\end{figure}

\begin{figure}[ht]
\begin{center}
\epsfig{file=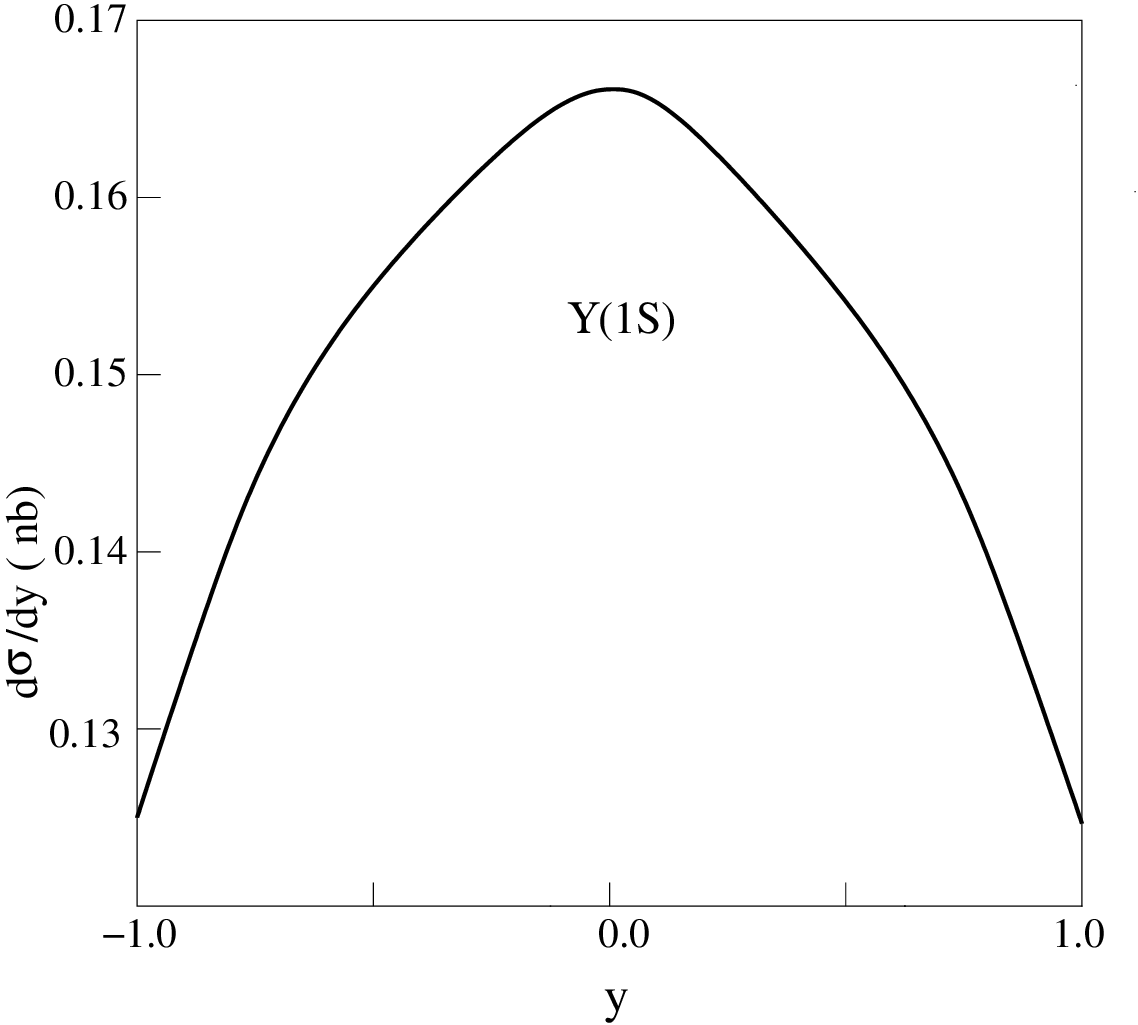,height=7cm,width=10cm}
\caption{d$\sigma$/dy for 2$m_b$=10 GeV, $\sqrt{s_{pp}}$=5.44 TeV via p-p 
collisions producing $\Upsilon(1S)$, $\lambda=0$}  
\end{center}
\end{figure}

\begin{figure}[ht]
\begin{center}
\epsfig{file=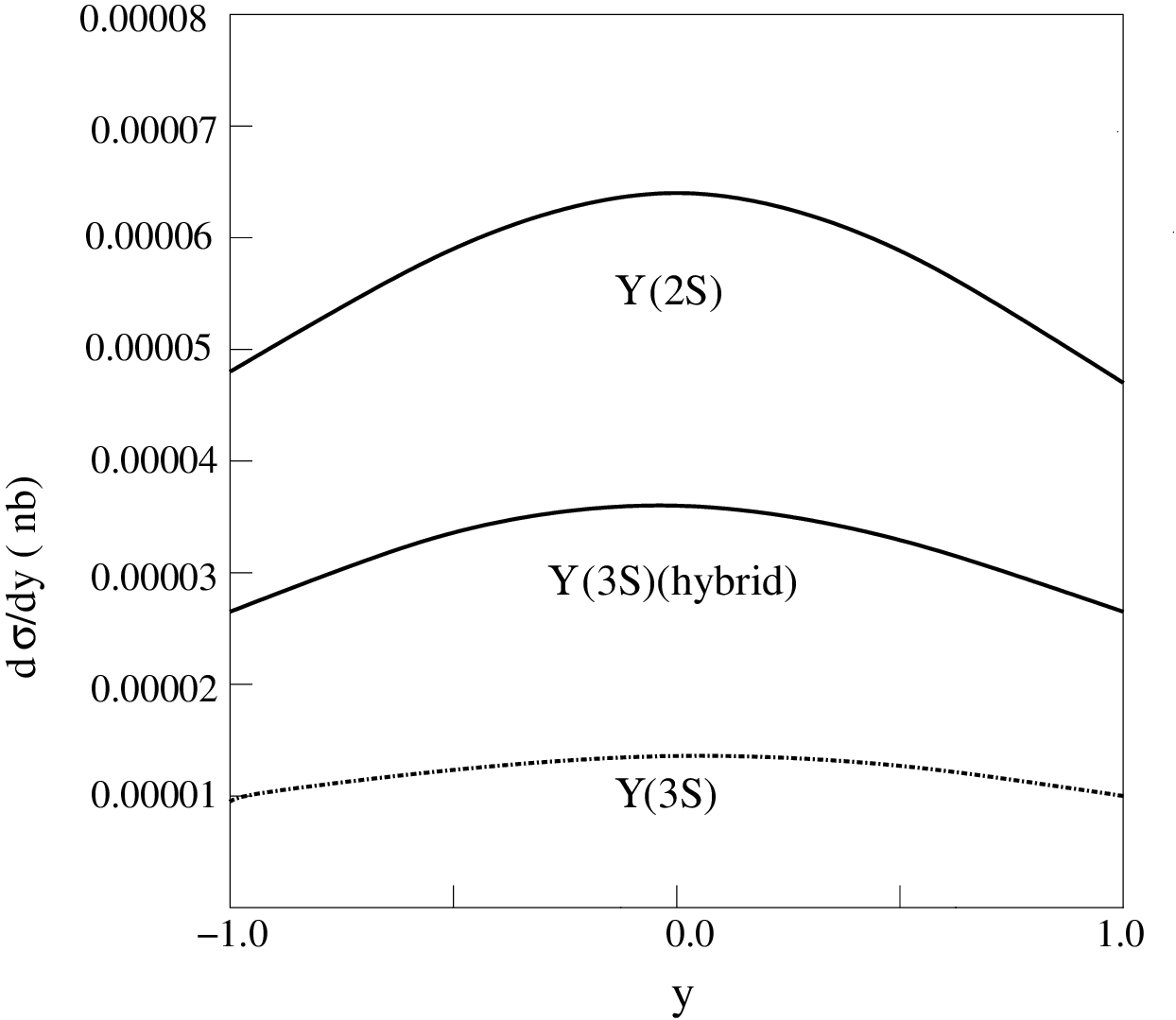,height=7cm,width=10cm}
\caption{d$\sigma$/dy for 2$m_b$=10 GeV, $\sqrt{s_{pp}}$=5.44 TeV via p-p
collisions producing with $\lambda=0$ $\Upsilon(2S)$ and $\Upsilon(3S)$(hybrid).
For $\Upsilon(3S)$ the dashed curve is for the standard $b\bar{b}$ model.}  
\end{center}
\end{figure}

\subsection{The differential rapidity cross section for the
 production of a heavy quark state $\Phi$ via O-O
  collisions}
\Large
\vspace{4mm}

Opportunities of O-O Collisions at the LHC and discussed in Ref~\cite{arxiv2104}.
Note that O represents the Oxygen nucleus with 8 protons. The most common Oxygen
nucleus has 8 protons, 8 neutrons and atomic number A=16 or
$O=O(p=8,n=8,A=16)$.For O-O collisions we use  $R^E_{OO}$ =0.25= $R^E_{XeXe}$/2.
\vspace{6mm}

The differential rapidity cross sections are shown in the following figures.
\vspace{6mm}

\begin{figure}[ht]
\begin{center}
\epsfig{file=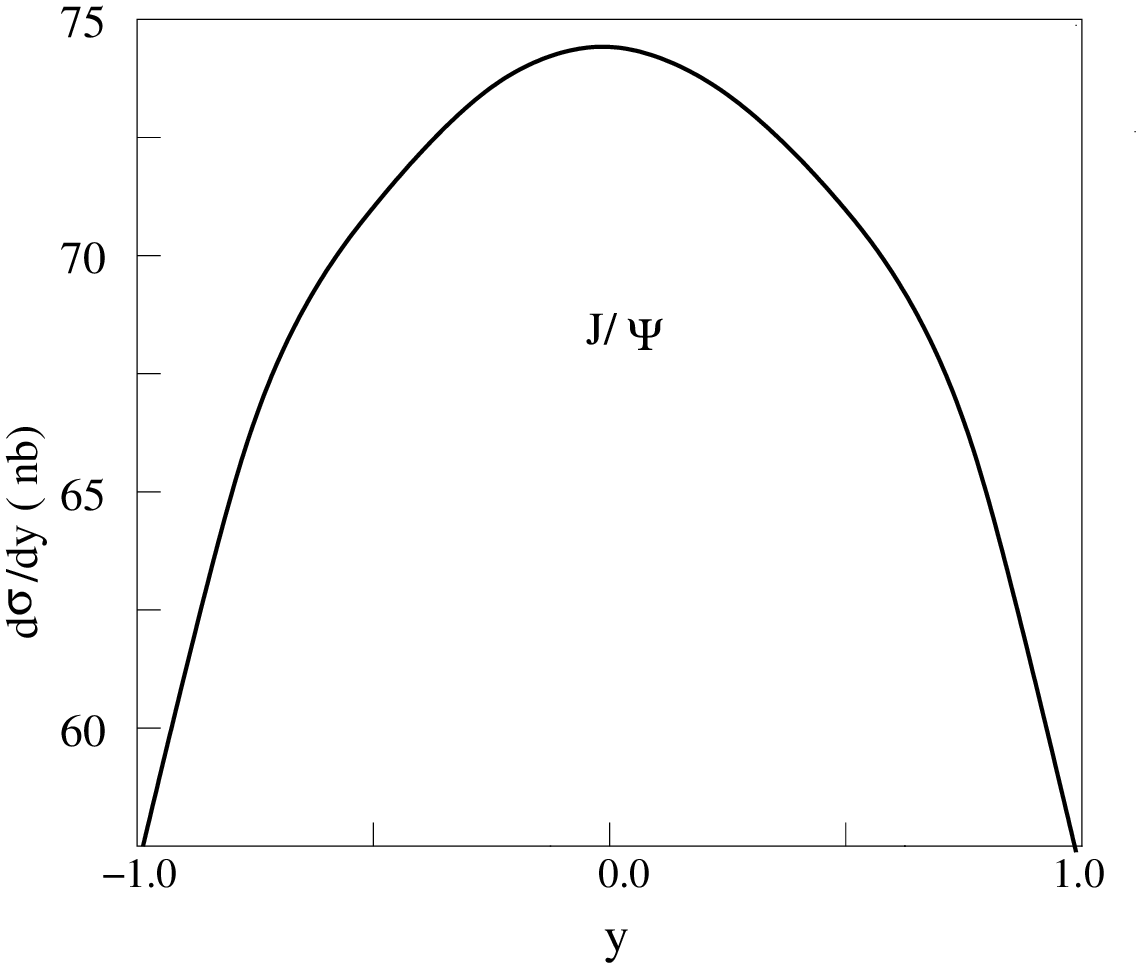,height=7cm,width=12cm}
\end{center}
\caption{d$\sigma$/dy for 2$m_c$=3 GeV, $\sqrt{s_{pp}}$=5.44 TeV via O-O
collisions producing $J/\Psi$ with $\lambda=0$}
\end{figure}

\begin{figure}[ht]
\begin{center}
\epsfig{file=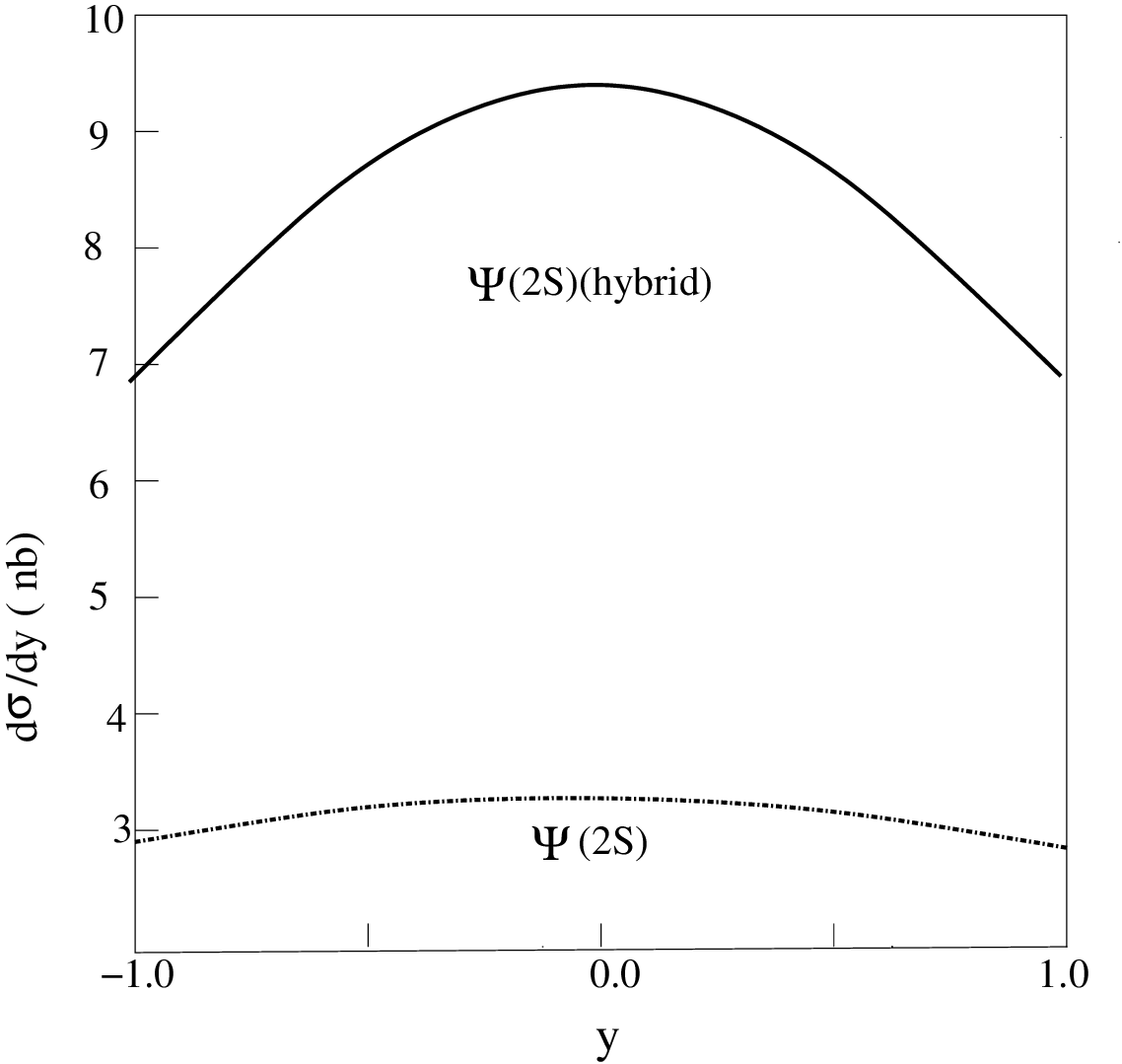,height=5cm,width=10cm}
\end{center}
\caption{d$\sigma$/dy for 2$m_c$=3 GeV, $\sqrt{s_{pp}}$=5.44 TeV via O-O
collisions producing $\Psi(2S)$, hybrid theory, with $\lambda=0$. The dashed 
curve is for the standard $c\bar{c}$ model.}
\end{figure}

\begin{figure}[ht]
\begin{center}
\epsfig{file=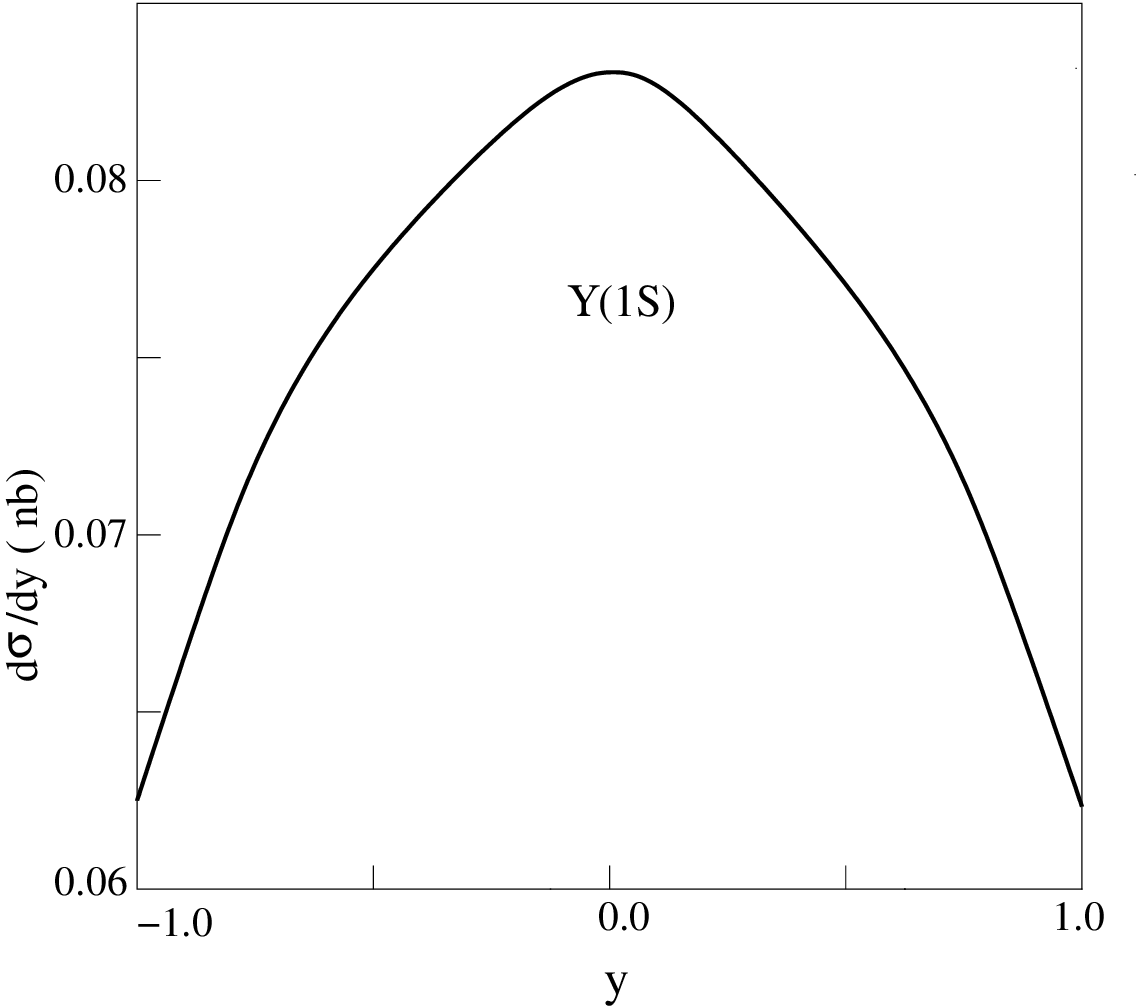,height=4cm,width=10cm}
\end{center}
\caption{d$\sigma$/dy for 2$m_b$=10 GeV, $\sqrt{s_{pp}}$=5.44 TeV via O-O
collisions producing $\Upsilon(1S)$, $\lambda=0$}
\end{figure}

\begin{figure}[ht]
\begin{center}
\epsfig{file=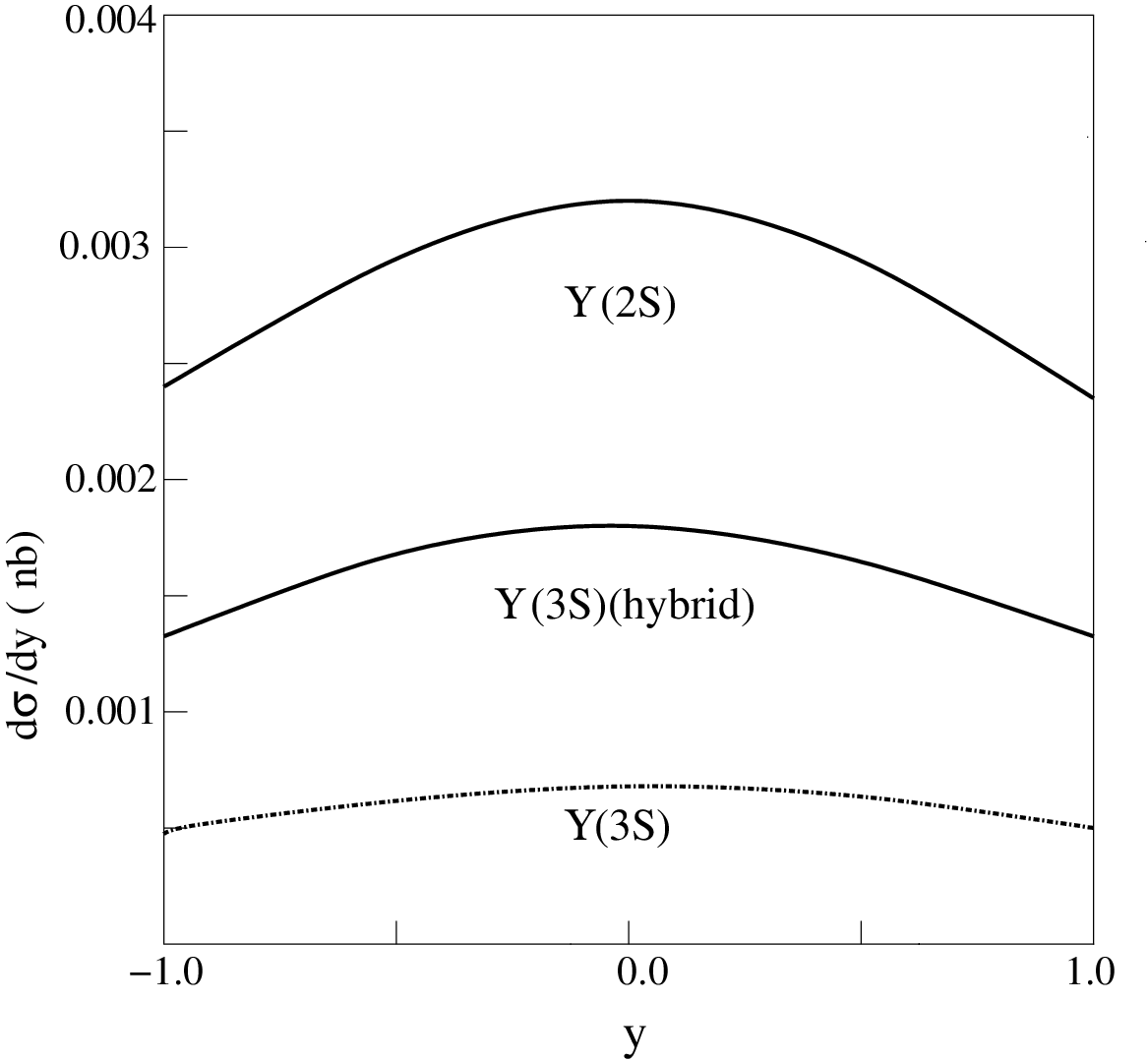,height=7cm,width=10cm}
\end{center}
\caption{d$\sigma$/dy for 2$m_b$=10 GeV, $\sqrt{s_{pp}}$=5.44 TeV via O-O 
collisions producing with $\lambda=0$ $\Upsilon(2S)$ and $\Upsilon(3S)$(hybrid).
For $\Upsilon(3S)$ the dashed curve is for the standard $b\bar{b}$ model.}
\end{figure}
 
Note that the differential rapidity cross section for the $\Upsilon(3S)$(hybrid)
is larger than the differential rapidity cross section for the  $\Upsilon(3S)$
and could be measured in future CERN LHC experiments .

\newpage
\section{Results and Conclusions}
\large

\vspace{3mm}

  In section 2, we review mixed heavy quark hybrid and p-p collisions.
In subsection 2.1, we discuss, $\Psi$ and $\Upsilon$ production via p-p
collisions with $\sqrt{s_{pp}}$ = 5.44 TeV. We give figures for the
differential rapidity cross sections for the production $\Psi(1S)$ and
hybrid $\Psi(2S)$ via p-p collisions.
\vspace{3mm}

In subsection 2.2 we discuss, $\Psi$ and $\Upsilon$ production via O-O collisions
with $\sqrt{s_{pp}}$ = 5.44 TeV. Also, we discuss the figures for O-O collisions
producing $J/\Psi$, hybrid $\Psi(2S)$, $\Upsilon(1S)$ and hybrid
$\Upsilon(3S)$.  It is noted in subsection 2.2, that the differential rapidity
cross section for the $\Upsilon(3S)$(hybrid), will be both a test of the
validity of the mixed hybrid theory and should also be a guide for LHC
experiments~\cite{ALICE:2022wpn}, for future measurements with small systems~\cite{Das:2023mye}.

\vspace{3mm}

{\bf Acknowledgements}

\vspace{3mm}

Author D.D. acknowledges the facilities of Saha Institute of Nuclear Physics, 
Kolkata, India. Author L.S.K. acknowledges support in part by a grant from
the Pittsburgh Foundation.


\begin{thebibliography}{99}
\bibitem{ALICE:2022wpn}
S.~Acharya \textit{et al.} [ALICE],Eur. Phys. J. C \textbf{84}, no.8, 813 (2024)
\bibitem{brahms05} I. Arsene et. al. (BRAHMS Collaboration) Nucl. Phys. A
{\bf 757},1 (2005)
\bibitem{kd17} Leonard S. Kisslinger and Debasish Das, arXiv:1612.02269/hep-ph;
JHEP 09, 105 (2017)
\bibitem{lsk09} L.S. Kisslinger, Phys. Rev. {\bf D 79}, 114026 (2009)
\bibitem{kd16} Leonard S. Kisslinger and Debasish Das, Int. J. Theor. Phys.
{\bf 55}, 5152 (2016)
\bibitem{klm14}L.S. Kisslinger, M.X. Liu, and P. McGaughey, Phys. Rev. 
{\bf C 89}, 024914 (2014)
\bibitem{cl96}P.L Cho and A.K. Leibovich, Phys. Rev. {\bf D 53}, 150 (1996)
\bibitem{bc96}E. Braaten and Y-Q Chen, Phys. Rev. {\bf D 54}, 3216 (1996)
\bibitem{fl96}E. Braaten and S. Fleming, Phys. Rev. Lett. {\bf 74}, 3327 (1995)
\bibitem{Das:2023mye}D.~Das,Curr. Sci. \textbf{125}, no.8, 820-821 (2023)
\bibitem{lsk16} Leonard S Kisslinger, Int. J. Theor. Phys. {\bf 545}, 1847 
(2016)
\bibitem{lskdd16} Leonard S Kisslinger and Debasish Das, Int.J.Mod.Phys.
{\bf A 13}, 1630010 (2016)
\bibitem{alice18}   S. Acharya $et\;al$ (ALICE collaboration), Phys. Lett. B
 {\bf 785}, 419 (2018)
\bibitem{cdf} CDF Collaboratin, Phys. Rev. Lett. {\bf 79}, 578 (1997)
\bibitem{vogel} H. Vogel, Proceedings of 4th Flavor Physics and CP Violation
Conference (FPCP'06) (2006)
\bibitem{klm11}L.S. Kisslinger, M.X. Liu, and P. McGaughey, Phys. Rev. 
{\bf D 84}, 114020 (2011)
\bibitem{star09} B.I. Abelev $et\;al$ (STAR Collaboration), Phys. Rev. 
{\bf C 80}, 041902 (2009)
\bibitem{star02} C. Adler $et\;al$ (STAR Collaboration), Phys. Rev. Lett.
{\bf 89}, 202301 (2002)
\bibitem{vitov06} I. Vitev, T. Goldman, M.B. Johnson, J. W. Qiu Phys. Rev. 
{\bf D 74}, 054010 (2006) 
\bibitem{sharma09} R. Sharma, I. Vitev, and B-W. Zhang, Phys. Rev. {\bf C 80},
054902 (2009) 
\bibitem{arxiv2104}https://arxiv.org/abs/2103.01939  

\end{thebibliography}
\end{document}